\documentclass[preprint,10pt,twocolumn,superscriptaddress,notitlepage,longbibliography]{revtex4-1}

\usepackage{array}
\usepackage{microtype}
\usepackage{natbib}

\usepackage{graphicx}
\usepackage{amssymb}
\usepackage{multirow}
\usepackage{color}
\usepackage{amsmath}
\usepackage{threeparttable}
\usepackage{subcaption}
\usepackage[nolist,nohyperlinks]{acronym}
\usepackage{hyperref}
\usepackage[nameinlink,capitalize]{cleveref} 
\usepackage{placeins}

\graphicspath{./figures/}

\begin{document}

\title{Stress Relaxation in Monodisperse Entangled Polymer Melts: Correlation Between Viscoelastic Response and Single-Chain Relaxation via Molecular Dynamics Simulations}

\author{Alireza F. Behbahani}
\email{aforooza@uni-mainz.de}
\affiliation{Institut f\"{u}r Physik, Johannes Gutenberg-Universit\"{a}t Mainz, Staudingerweg 7, D-55099 Mainz, Germany}

\begin{abstract}

We study stress relaxation in several types of entangled monodisperse linear polymer melts by comparing the shear stress relaxation modulus, $G(t)$, with the end-to-end vector autocorrelation function, $P(t)$. The study includes three Kremer-Grest bead-spring models with varying chain stiffness, as well as a chemistry-specific coarse-grained model of \emph{cis}-1,4-polybutadiene. For each model, multiple chain lengths were simulated, spanning a range of $N/N_e = 5$-$50$ entanglements per chain.
We observe that in all cases the behavior of $G(t)$, beyond the short-time Rouse regime, is accurately described by $G^0_{\mathrm{N}}[P(t)]^2$, where the chain-length-independent prefactor $G^0_{\mathrm{N}}$ denotes the plateau modulus.
This correlation is consistent with both double reptation and dynamic tube dilation models of polymer relaxation, although the two models are based on different physical pictures.
The double reptation model represents the melt as a transient network in which stress relaxation is governed by the survival probability of pairwise entanglements. The dynamic tube dilation model, however, assumes that the tube of constraints surrounding a polymer chain progressively enlarges as relaxation proceeds.
The relation $G(t) = G^0_\mathrm{N}[P(t)]^2$ can serve as a basis for determining the plateau modulus and the corresponding entanglement length. It also simplifies the modeling of $G(t)$, since an accurate analytical expression for $P(t)$ is sufficient to describe the long-time behavior of $G(t)$. We further compare the simulation data for $P(t)$ and $G(t)$ with theoretical predictions.

\end{abstract}
 
 \maketitle
 
\section{Introduction}   
\label{sec:intro}

The investigation of the viscoelastic properties of polymers has been an important research area since the emergence of polymer science~\cite{ferry1980viscoelastic}.
This interest has been driven in part by the practical importance of the understanding of the viscoelastic behavior of polymers. Furthermore, the viscoelastic properties have been studied to gain insight into the relaxation processes of polymers and to test fundamental theories of polymer dynamics.

In a polymer melt, a chain cannot cross its surrounding chains. This topological constraint profoundly affects the dynamics of long chains, preventing them from undergoing unrestricted Brownian motion, as assumed in the Rouse model.
The tube model represents the restraining influence of surrounding chains as a confining tube along the chain~\cite{de1971reptation,doi1988theory}. 
To renew its conformation, the chain must escape the tube (formed at time $t = 0$) by moving along the tube contour. While moving along the tube contour, the chain initially undergoes incoherent restricted Rouse motion. Then, after its Rouse time, when it has traveled a distance along the tube comparable to its end-to-end distance, the motion becomes coherent curvilinear diffusion along the tube. This curvilinear diffusion is known as reptation~\cite{rubinstein2003polymer,mcleish2002tube}.
For very long chains, the contour length of the tube is much larger than the end-to-end distance of the chain, 
 and the chain performs reptation along most of its path within the tube.
Therefore, in the basic tube model, it is assumed that the chain moves along  the tube only by reptation, and the contribution of incoherent motion in the escape of the chain is ignored~\cite{doi1988theory}.
This assumption is reasonable in the limit of very long chains; however, over a wide range of chain lengths,
the contribution of incoherent restricted Rouse motion along the tube cannot be neglected.
This incoherent motion of the chain along the tube allows part of the chain to exit the tube at short times, and reduces the effective distance the chain must travel by reptation to escape the tube.
This motion is usually viewed as a process that leads to fluctuations in the contour length of the tube, since at times shorter than the Rouse time, the independent motion of the two chain ends induces variations in the contour length. The process is also referred to as contour length fluctuations (CLF)~\cite{doi1983explanation,mcleish2002tube}.

The combination of reptation and contour-length fluctuations (CLF), which together describe the motion of a chain within a fixed tube, explains several key features of entangled polymer dynamics. These include the chain-length dependencies of the self-diffusion coefficient and longest relaxation time,
as well as characteristic behaviors of dynamical observables such as the mean-squared displacement, end-to-end vector correlation, and single-chain dynamic structure factor for sufficiently long chains~\cite{doi1983explanation,mcleish2002tube}.
This fixed-tube picture, however, does not adequately describe the viscoelastic spectrum of even very long monodisperse linear polymer melts~\cite{rubinstein1988self,rubinstein2003polymer,mcleish2002tube}.

According to the fixed-tube model, at long times, stress relaxation occurs only through the escape of the polymer chain from its confining tube. Consequently, the stress relaxation modulus, $G(t)$, is expected to be proportional to the \emph{surviving tube fraction}, $\mu(t)$, such that 
$G(t) = G^0_\text{N} \mu(t)$ where $G^0_\text{N}$ is the plateau modulus, which is independent of chain length, and $\mu(t)$ represents the fraction of the chain that remains confined within its original tube formed at $t = 0$~\cite{doi1988theory,mcleish2002tube}.
However,  this relation, when the analytical expression for $\mu(t)$ is used, does not yield a satisfactory description of the viscoelastic spectrum~\cite{des1990relaxation,van2002evaluation}.

Within the framework of the tube model, the inadequacy of the fixed-tube picture in describing the viscoelastic spectrum is attributed to the process of constraint release (CR). This process arises from the motion of the surrounding chains that constitute the confining tube of a given (or target) chain.
CR process has been viewed in different ways in the literature. 
According to one approach, a CR event caused by the motion of a surrounding chain is assumed to result in a local displacement of the tube.
These local displacements due to CR events lead to a slow, Rouse-like motion of the tube, which in turn contributes to stress relaxation~\cite{rubinstein1988self,rubinstein2003polymer}.
Rubinstein and Colby developed a theory based on the above “Rouse-tube” description of the CR process~\cite{rubinstein1988self}. They assumed that the tube could be modeled as an effective Rouse chain with random bead mobilities, whose distribution depends on $\mu(t)$. 
It was also assumed that the stress relaxation modulus of a monodisperse melt can be written as $G(t) = G^0_\text{N}\mu(t)R(t)$ where $R(t)$ is the stress relaxation modulus of the effective Rouse-like chains that mimic the tube motion induced by CR.
The Rubinstein–Colby model of CR has also been incorporated into the Likhtman–McLeish~\cite{likhtman2002quantitative} model for the viscoelasticity of entangled melts. The adjustment in the latter model is the addition of a fitting parameter in the $R(t)$ function, which allows control over its decay rate. The resulting expression for the relaxation modulus is $G(t) = \mu(t)R(t,c_\nu)$, where $c_\nu$ is a fitting parameter.  The limit $c_\nu = 0$ corresponds to neglecting constraint release, for which $R(t) = 1$, while $c_\nu = 1$ recovers the original Rubinstein–Colby model.

Besides the aforementioned model, the CR process has been described as a mechanism that causes an increase in the effective tube diameter over time (i.e., dynamic tube dilation). This model was first introduced by Marrucci~\cite{marrucci1985relaxation} and later adopted by other researchers~\cite{matsumiya2000comparison,watanabe2001dielectric,watanabe2009slow}. Marrucci assumed that the portion of the chain that escapes from the tube behaves as a solvent, which leads to an increase in the tube diameter. For a monodisperse melt, this model yields $G(t) = G^0_\text{N} [\mu(t)]^{1+\alpha}$ with $\alpha \approx 1$.

A similar relationship was proposed by des Cloizeaux for monodisperse linear polymer melts; however, it was introduced based on a much simpler physical picture, which was called \emph{double reptation}~\cite{des1988double,des1990relaxation,des1993polymer}.  
The main assumptions of double reptation are 
(i) An entanglement between two chains produces a stress point;
(ii) the stress relaxation modulus is proportional to the number of original stress points that remain at time $t$; 
and (iii) a stress point persists if neither the first chain nor the second chain reptates through that point. The survival probability of a stress point is $[\mu(t)]^2$, and hence the stress relaxation modulus is given by $G(t) = G^0_\text{N} [\mu(t)]^2$.
The exponent $2$ arises because a stress point is produced by an entanglement between two chains.
In other words, in the double reptation description, a polymer melt is viewed as a transient network formed by entanglements. The nodes of this network are created by entanglements between pairs of chains, and the stress relaxation modulus reflects the survival probability of the original stress points.

The Rouse–tube model of constraint release (adopted in the Likhtman–McLeish model~\cite{likhtman2002quantitative})
and the double reptation model have been employed to describe viscoelastic spectra of monodisperse linear polymer melts obtained from experiments~\cite{rubinstein1988self,van2002evaluation,likhtman2002quantitative,auhl2008linear,glomann2011unified} as well as from molecular simulation techniques~\cite{hou2010stress,li2021dynamics,ghanta2025modeling}.
To apply these models, one needs an analytical expression for $\mu(t)$ which is then used either in the Rouse–tube model of constraint release or in the relation $G(t) = G^0_\text{N} [\mu(t)]^2$. The resulting analytical expressions for $G(t)$ take as inputs physical quantities such as the entanglement length and entanglement time, as well as additional fitting parameters introduced in an ad hoc manner. 
Therefore, comparison of these analytical relations with experimental data usually corresponds to evaluating whether the models can adequately fit the experimental results.

A more direct insight into the viscoelastic response can be obtained by comparing the shear stress relaxation modulus, $G(t)$, with the end-to-end vector autocorrelation function, $P(t)$.
In the tube model, $P(t)$, is equivalent to the surviving tube fraction~\cite{mcleish2002tube}. 
Consequently, the fixed tube model leads to the prediction $G(t) = G^0_\text{N} P(t)$, while the predictions of the double reptation and dynamic tube dilation models can be written as $G(t) = G^0_\mathrm{N}[P(t)]^2$  (we refer to this relation as the double reptation relation, although it also arises from the dynamic tube dilation model).
However, the latter relation does not arise from the Likhtman–McLeish model. As mentioned above, in this model stress relaxation modulus is given by $G(t) = G^0_\text{N}\mu(t)R(t,c_\nu)$. Depending on the value of $c_\nu$, the function $R(t,c_\nu)$ can differ significantly from $\mu(t)$, and as a result the quantity $G^0_\text{N}\mu(t)R(t,c_\nu)$ differs from $G^0_\text{N}[\mu(t)]^2$ (or equivalently $G^0_\text{N}[P(t)]^2$).
Even with $c_\nu = 1$, which corresponds to the original Rouse-tube model of Rubinstein and Colby, $R(t,c_\nu)$ exhibits a long-time tail and deviates from $\mu(t)$~\cite{likhtman2002quantitative}.

The end-to-end vector autocorrelation function, $P(t)$, is experimentally accessible via dielectric relaxation spectroscopy of type-A polymers, which have dipole moments aligned along the polymer backbone~\cite{watanabe2001dielectric,mcleish2002tube}. The proportionality between $G(t)$ and $[P(t)]^2$ has been examined through combined rheological and dielectric relaxation spectroscopy measurements for monodisperse \emph{cis}-polyisoprene chains (with $N/N_\text{e} = 10\text{--}30$, where $N_\text{e}$ is the entanglement length). Consistent with the predictions of the double reptation and dynamic tube dilation models, the relation was reported to be approximately valid~\cite{matsumiya2000comparison}.
It is worth noting that combined rheological and dielectric relaxation spectroscopy measurements have also been used to test the predictions of the tube dilation model in mixtures of short and long polymer chains~\cite{watanabe2004viscoelastic,watanabe2004test,van2012effective}. However, for such systems, it has been reported that the prediction of dynamic tube dilation, in its basic form, is not generally valid. 
This limitation has been attributed to the rapid relaxation of the short chains, which occurs on a timescale too short for the long chains to adjust their motion and fully explore their dilated tube~\cite{watanabe2004test,van2012effective}.

Few computational studies have also investigated the validity of the relation
$G(t) = G^0_\text{N} [P(t)]^2$
for monodisperse polymer melts. This relation was observed using a multi-chain slip-link model (for $N/N_\text{e} = 10$)~\cite{masubuchi2001brownian}. More recently, we~\cite{behbahani2024relaxation} have also provided support for this relation through simulations of the standard fully flexible Kremer–Grest model. Using this model, we simulated chain lengths up to $N/N_\text{e} = 20$, although good statistics for $G(t)$ at long times were obtained  up to $N/N_\text{e} = 8$. For the model studied, we found that $G^0_\text{N}[P(t)]^2$ accurately describes the long-time behavior of $G(t)$~\cite{behbahani2021dynamics}.

In the present contribution, we provide further support for the relation $G(t) = G^0_\text{N}[P(t)]^2$ by studying several types of entangled, monodisperse linear polymer melts using molecular dynamics simulations over a significantly broader range of chain lengths, characterized by the number of entanglements per chain. The study includes both generic and chemistry-specific models. We also use the relation $G(t) = G^0_\text{N}[P(t)]^2$ as a basis for determining the plateau modulus and its corresponding entanglement length. Furthermore, within this framework, an accurate analytical expression for $P(t)$ is sufficient to predict $G(t)$, excluding its short-time behavior. We therefore compare the $P(t)$ and $G(t)$ data obtained from simulations with theoretical predictions.

We study three Kremer-Grest bead-spring models with varying chain stiffness~\cite{kremer1990dynamics,auhl2003equilibration,everaers2020kremer}, as well as a chemistry-specific coarse-grained model parameterized to represent \emph{cis}-1,4-polybutadiene~\cite{behbahani2021dynamics}. For each model, chains of different lengths are considered.
The Kremer–Grest bead–spring models with different stiffness values have recently been employed to represent a wide range of commodity polymers~\cite{everaers2020kremer}.
Furthermore, the entanglement length, $N_\text{e}$, decreases with increasing chain stiffness, and the stiffer bead–spring models enable simulations of chains with larger $N/N_\text{e}$ ratios.
In the present study, chain lengths corresponding to $N/N_\text{e} = 5$–$50$ are considered.
For the bead–spring models, the $N_\text{e}$ values obtained from the double reptation relation are compared with those determined from primitive path analysis, which is a widely used method for determining entanglement length and for characterizing entanglement networks in molecular dynamics simulations~\cite{everaers2004rheology,hoy2009topological,moreira2015direct}.

The manuscript is organized as follows. In \cref{sec:model}, we introduce the models under study and provide the details of the simulations. In \cref{sec:results}, we examine the correlation between $G(t)$ and $P(t)$ and compare the simulation results with theoretical predictions. Finally, in \cref{sec:summary}, we summarize the main findings.

\section{Model}
\label{sec:model}

Molecular dynamics simulations were conducted using three Kremer–Grest bead–spring models of varying stiffness~\cite{kremer1990dynamics,auhl2003equilibration}, as well as a chemistry-specific model that represents \emph{cis}-1,4-polybutadiene (\emph{cis}-PB)~\cite{behbahani2021dynamics}.

In the Kremer-Grest models, all beads have mass $m$ and interact via purely repulsive nonbonded interactions described by the Weeks-Chandler-Andersen potential, characterized by a length scale $\sigma$ and energy scale $\varepsilon$~\cite{kremer1990dynamics}. Bonded interactions are modeled using a FENE
potential, which prevents chain crossing.
All quantities are expressed in units of $\sigma$, $\varepsilon$, and $m$. The bead number density is $\rho = 0.85\,\sigma^{-3}$. The temperature was maintained at $k_\mathrm{B}T = 1\, \varepsilon$ using a Langevin 
thermostat with friction coefficient $\gamma = 0.5\, \tau^{-1}$. A time step of $0.01\, \tau$ was used.
The stiffness of the chains is controlled by an angle-bending potential of the form $U(\theta) = k_{\theta}(1 - \cos\theta)$, where $\theta$ is the angle between the vectors representing two consecutive bonds~\cite{auhl2003equilibration}.
We studied three models with $k_{\theta} = 1.5$, $k_{\theta} = 2.5$, and $k_{\theta} = -1$. 
Larger values of $k_{\theta}$ correspond to stiffer chains with larger end-to-end distances.
Furthermore, increasing $k_{\theta}$ leads to a decrease in the entanglement length, $N_\text{e}$. 
Using primitive path analysis, the estimated $N_\text{e}$ values are $28$, $15$, and $120$ beads for $k_{\theta} = 1.5$, $2.5$, and $-1$, respectively~\cite{hsu2016static,everaers2020kremer}. In the next section, we compare these estimates with the $N_\text{e}$ values obtained from the viscoelastic spectra.

\begin{figure}[htb!]
    \centering
        \includegraphics[width=0.45\textwidth]{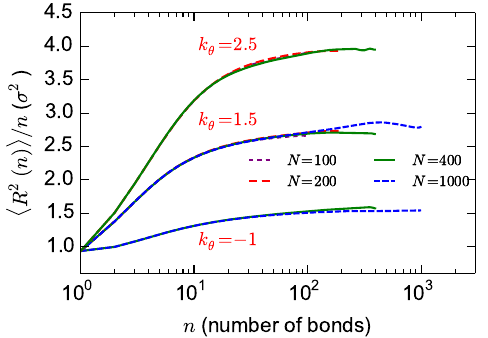}
\caption{Mean-squared internal distances for the simulated Kremer-Grest chains. $R^2(n)$ denotes the average squared distance between two monomers separated by $n$ consecutive bonds along one chain.}
    \label{Fig:int-dist} 
\end{figure}

For each value of $k_{\theta}$, we simulated polymer chains of different lengths. For $k_{\theta} = 1.5$, chain lengths $N = 100$, $200$, $400$, and $1000$ were examined. For $k_{\theta} = 2.5$, results are reported for chain lengths $N = 100$, $200$, and $400$. Finally, for $k_{\theta} = -1$, simulations were performed for $N = 400$ and $1000$.
For chain lengths $N = 100$, $200$, $400$, and $1000$, the simulation box contained $800$, $400$, $384$ (or $96$), and $216$ chains, respectively.
Production runs were performed for approximately $2 \times 10^7 \tau$ to $2 \times 10^8 \tau$ ($2\times 10^9$ to $2 \times 10^{10}$ time steps).
The initial pre-equilibrated configurations for the bead-spring simulations were taken from our previous simulations of the standard Kremer–Grest model, in which the angle-bending potential was not applied (corresponding to $k_{\theta} = 0$)~\cite{behbahani2024relaxation}.
For equilibration, the initial configurations were subjected to long equilibration runs. The equilibration of the resulting configurations was verified by measuring the internal chain distances, as shown in \cref{Fig:int-dist}. This figure presents $R^2(n)/n$ as a function of $n$, where $R^2(n)$ represents the mean-squared distance between two monomers separated by $n$ bonds along a chain. At large length scales, the internal distances approximately follow the expected Gaussian behavior ($R^2(n)\sim n$).
For large $n$, the ratio $\langle R^2(n) \rangle / n$ approaches the statistical segment length squared $b^2$, such that $\langle R^2(n) \rangle = n b^2$. For models with stiffness parameters $k_\theta = -1$, $1.5$, and $2.5$, the statistical segment length $b$ is calculated to be $1.25\,\sigma$, $1.65\,\sigma$, and $1.99\,\sigma$, respectively.

The chemistry-specific coarse-grained (CG) model of \emph{cis}-PB was parametrized based on a united-atom model~\cite{behbahani2021dynamics}. In the CG model, each monomer of \emph{cis}-PB, containing four CH$_x$ groups, is mapped onto a single CG bead. The model includes non-bonded interactions, as well as bond-stretching, angle-bending, and dihedral potentials, which are derived by matching the local structural distributions of the CG model to those of the atomistic model using the iterative Boltzmann inversion method.
After proper time scaling, this model faithfully reproduces the dynamical properties of the atomistic model for time scales larger than the segmental relaxation time and length scales greater than the CG bead size.
This CG model significantly extends the accessible simulation time scale relative to the atomistic model. Specifically, it enables simulations of \emph{cis}-PB for times on the order of $100\,\mu\mathrm{s}$ at $413\,\mathrm{K}$.
More details about the CG model can be found elsewhere~\cite{behbahani2021dynamics}.
In the current work, we study \emph{cis}-PB chains with lengths $N = 200$, $400$, and $1000$ monomers at $T = 413$ K. For $N = 400$ and $1000$, the data for the stress relaxation modulus, $G(t)$, are identical to those presented in the previous study~\cite{behbahani2021dynamics}. However, in the present work, the $G(t)$ data for $N = 200$ are slightly different, as we extended the run to improve the statistics.

The simulations of the Kremer–Grest models were performed using the LAMMPS~\cite{thompson2022lammps} and HOOMD-blue~\cite{anderson2020hoomd} packages, while the simulations of the chemistry-specific CG model were carried out using the GROMACS package~\cite{pronk2013gromacs}.

\section{Results}
\label{sec:results}

\subsection{Correlation between viscoelastic response and single-chain dynamics}

\label{sec:double-rep}

In this section, we examine the validity of the following relation, as proposed by the double reptation model and also by the dynamic tube dilation model, for the polymer melts studied in our simulations:
\begin{equation}
 G(t) = G^0_\text{N}[P(t)]^2.
 \label{Eq:doubel-rep}
\end{equation}
As noted in Section~\ref{sec:intro}, $G(t)$ is the shear stress 
relaxation modulus, $P(t)$ is the end-to-end vector autocorrelation function, and $G^0_\text{N}$ is the plateau modulus. The latter is related to the entanglement length via~\cite{fetters1994connection,likhtman2002quantitative} 
\begin{equation}
G^0_\text{N} = \frac{4}{5} \frac{\rho k_\text{B}T}{N_\text{e}},    
\end{equation}
where $\rho$ is the monomer number density, $k_\text{B}$ is the Boltzmann constant, and $T$ is the temperature. 
The factor $4/5$, which is absent from the corresponding relation for crosslinked rubbers, originates from the possibility of monomer redistribution along the chain in a deformed entangled melt (i.e., relaxation of longitudinal modes)~\cite{doi1988theory,likhtman2002quantitative,mcleish2002tube,masubuchi2020entanglement}.
In equilibrium simulations, $G(t)$, is calculated from the autocorrelation function of shear stresses~\cite{likhtman2007linear,behbahani2024local}:
\begin{equation}
    G(t) = \frac{V}{k_\text{B}T}
    \langle \sigma_{\alpha \beta}(t) 
    \sigma_{\alpha \beta}(0)\rangle,\ \ \  \alpha \neq \beta.
    \label{Eq:Gt}
\end{equation}
Here, $V$ denotes the volume, $\sigma_{\alpha \beta}(t)$ ($\alpha, \beta \in \{x, y, z\}$) represents a shear component of the stress tensor, and $\langle \cdot \rangle$ denotes the ensemble average. The correlation function was computed using the multiple-$\tau$ correlator algorithm, and to improve statistics, results were averaged over different orientations of the coordinate system~\cite{ramirez2010efficient}.
The end-to-end vector autocorrelation function, $P(t)$, is defined as:
\begin{equation}
    P(t) = \frac{\langle {\mathbf{R}}(t). {\mathbf{R}}(0)\rangle}{\langle {\mathbf{R}}(0). {\mathbf{R}}(0)\rangle},
\end{equation}
where ${\mathbf{R}}(t)$ is the end-to-end vector at time $t$.

To examine the validity of \cref{Eq:doubel-rep}, we investigate whether a chain-length-independent prefactor, $G^0_\text{N}$, can be identified that satisfies this relation.

\begin{figure*}[htb!]
    \begin{subfigure}{0.45\textwidth} 
        \includegraphics[width=1.0\textwidth]{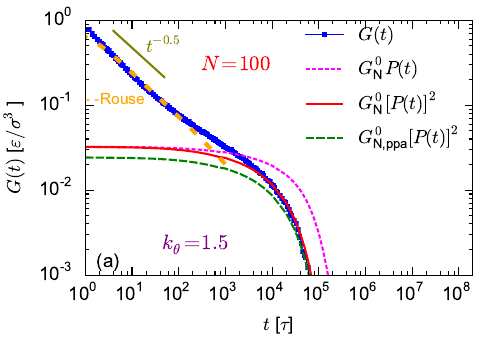}
    \end{subfigure}
    \begin{subfigure}{0.45\textwidth} 
        \includegraphics[width=1.0\textwidth]{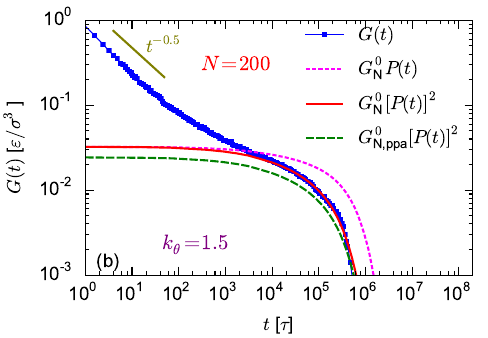}
    \end{subfigure}

     \begin{subfigure}{0.45\textwidth} 
        \includegraphics[width=1.0\textwidth]{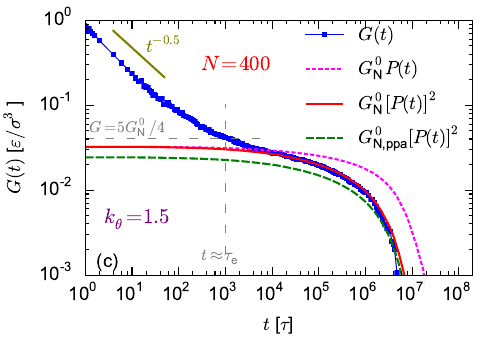}
    \end{subfigure}
        \begin{subfigure}{0.45\textwidth} 
        \includegraphics[width=1.0\textwidth]{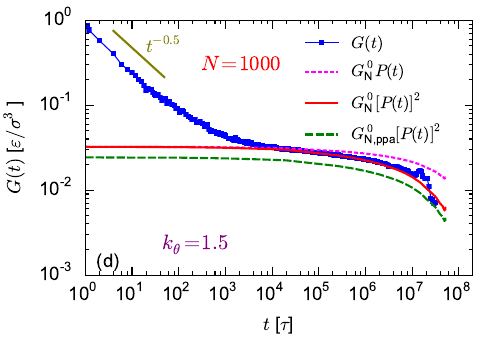}
    \end{subfigure}
\caption{The $G(t)$ curves for the bead–spring chains with a bending stiffness of $k_{\theta} = 1.5$. The $G(t)$ curves are compared to $G^0_\text{N}[P(t)]$ (dashed magenta), $G^0_\text{N}[P(t)]^2$ (solid red), and $G^0_\text{N,ppa}[P(t)]^2$ (dashed green), with $G^0_\text{N} = 0.0324\,\varepsilon/\sigma^3$ (corresponding to $N_\text{e} = 21$ beads) and $G^0_\text{N,ppa} = 0.0243,\varepsilon/\sigma^3$ (corresponding to $N_\text{e,ppa} = 28$ as calculated from primitive path analysis).
Panels correspond to chain lengths $N = 100$ (a), $200$ (b), $400$ (c), and $1000$ (d).
The dashed orange lines in panel (a) also show the short-time prediction of the Rouse model with monomeric time $\tau_0 = 2.1\,\tau$. 
Additionally, panel (c) shows an estimate of the entanglement time, obtained from the condition $G(\tau_\text{e}) = \rho k_\text{B} T / N_\text{e} = 5 G^0_\text{N}/4$.}
    \label{Fig:Gt-Pt-K1.5} 
\end{figure*}

\cref{Fig:Gt-Pt-K1.5} presents the measured $G(t)$ curves for the semi-flexible bead–spring chains with a bending stiffness of $k_{\theta} = 1.5$.  
The panels of this figure show $G(t)$ for chain lengths $N = 100$, $200$, $400$, and $1000$, separately.
The $G(t)$ curves exhibit a short-time Rouse regime, with $G(t) \sim t^{-1/2}$, before the chains feel entanglement constraints~\cite{doi1988theory,likhtman2007linear}.  
After this initial Rouse regime, the behavior of $G(t)$ is governed by entanglements.
The results in \cref{Fig:Gt-Pt-K1.5} show that, beyond the initial Rouse regime, the behavior of $G(t)$ for different chain lengths is well described by $G^0_\text{N}[P(t)]^2$, with $G^0_\mathrm{N} = 0.0324\,\varepsilon/\sigma^3$, corresponding to an entanglement length  $N_\text{e} = 21$ beads.
In all cases the $G(t)$ curve decays significantly faster than $G^0_\text{N}P(t)$, which represents the prediction of the fixed-tube model (i.e., ignoring CR). 
Furthermore, the long-time behavior of the $G(t)$ curves cannot be accurately captured if $G^0_{\text{N,ppa}} = 0.0243\, \varepsilon/\sigma^3$, corresponding to $N_\text{e,ppa} = 28$~\cite{hsu2016static} as calculated from primitive path analysis, is used as the prefactor of $[P(t)]^2$.
Since $G^0_{\text{N,ppa}}$ is lower than the estimated value for the plateau modulus $G^0_\text{N}$, the expression $G^0_\text{N,ppa}[P(t)]^2$ slightly underestimates the height of the late-time shoulder of the measured $G(t)$ curves.

As an additional analysis of the above values for the entanglement length and the corresponding plateau modulus, we
perform a self-consistency check, following the procedure used in our previous work~\cite{behbahani2024relaxation}. For a given value of $N_\text{e}$ (or its corresponding $G^0_\text{N}$), one can estimate entanglement time, $\tau_\text{e}$ from the $G(t)$ curves using $G(\tau_\text{e}) = \rho k_\text{B}T/N_\text{e} =  5 G^0_\text{N}/4$~\cite{likhtman2002quantitative,mcleish2002tube} (see \cref{Fig:Gt-Pt-K1.5}c). This value of $\tau_\text{e}$ is expected to be consistent with the Rouse time of a subchain with length $N_\text{e}$, namely $\tau_0 N_\text{e}^2$, where $\tau_0 = \zeta b^2 /(3\pi^2 k_\text{B}T)$ is the monomeric time~\cite{doi1988theory}. Here, $\zeta$ is the monomeric friction coefficient and $b$ is the statistical segment length.

We determine $\tau_0$ by matching the short-time Rouse regime of the simulated $G(t)$ curves to the prediction of the Rouse model~\cite{likhtman2012chapter}, which contains $\tau_0$ as its only input parameter ($G(t) = (\rho k_\text{B} T/N) \sum_p \exp[-2p^2t/(\tau_0 N^2)]$).
For the Kremer–Grest model with $ k_\theta = 1.5 $, this procedure yields $\tau_0 = (2.1 \pm 0.3)\,\tau $. The corresponding Rouse-model prediction is shown in panel (a) of \cref{Fig:Gt-Pt-K1.5} as orange dashed lines.
For this model, the statistical segment length is $b = 1.65\,\sigma$ (as noted in \cref{sec:model}), and the obtained value for $\tau_0$ 
therefore corresponds to a monomeric friction coefficient of $\zeta = 23\, m/\tau$. This value is comparable to a previously reported estimate $\zeta = (27 \pm 20\%)\, m/\tau$ based on mean-squared displacements~\cite{svaneborg2020characteristic}.

With $\tau_0 = 2.1\, \tau$, we carry out the self-consistency check described above.
Based on the $G(t)$ curves (for $N = 400$ and $1000$), taking $N_\mathrm{e} = 21$ gives an entanglement time $\tau_\mathrm{e} \approx 1000\,\tau$ (see \cref{Fig:Gt-Pt-K1.5}c). This value is consistent with the Rouse time of a subchain of length $N_\mathrm{e} = 21$, namely $\tau_0 N_\mathrm{e}^2 = 926\,\tau$.
In contrast, using $N_{\mathrm{e,ppa}} = 28$ yields $\tau_\mathrm{e} \approx 5200\,\tau$, which is much larger than the corresponding Rouse time of a subchain of length $N_{\mathrm{e,ppa}}$, $\tau_0 N_{\mathrm{e,ppa}}^2 = 1650\,\tau$.
Therefore, the value $N_\mathrm{e} = 21$ obtained from the double-reptation relation better satisfies this self-consistency check.

Hsu and Kremer~\cite{hsu2016static} previously measured $G(t)$ for the Kremer–Grest model with $k_\theta = 1.5$ for chains with lengths $N = 500$, $1000$, and $2000$. 
They reported that $N_{\text{e,ppa}} = 28$ is fully consistent with the plateau modulus extracted from the measured $G(t)$ curves. However, this conclusion is not fully supported by the results shown in \cref{Eq:doubel-rep}, which suggest that $N_{\text{e,ppa}}$ slightly underestimates the plateau modulus.
Their conclusion was based on visual inspection of $G(t)$ curves obtained from relatively short trajectories (over ten times shorter than the trajectories used in the present work, made possible by advances in computational power), which may introduce some uncertainty in their estimate, particularity comparable to the relatively small difference between $N_{\text{e,ppa}}$ and $N_{\text{e}}$ extracted from the double-reptation analysis.
Likhtman et al.~\cite{likhtman2007linear} also reported that, for bead–spring chains of varying stiffness (induced by an angle-bending potential different from the one used in this work), the plateau modulus predicted by primitive path analysis is systematically lower than the plateau modulus determined from an equivalent slip-spring model.
For the standard Kremer–Grest model with $k_\theta = 0$, our previous results similarly suggested that the entanglement length obtained from primitive path analysis underestimates the plateau modulus~\cite{behbahani2024relaxation}.

\begin{figure}[htb!]
    \centering

\begin{subfigure}{0.45\textwidth} 
        \includegraphics[width=1.0\textwidth]{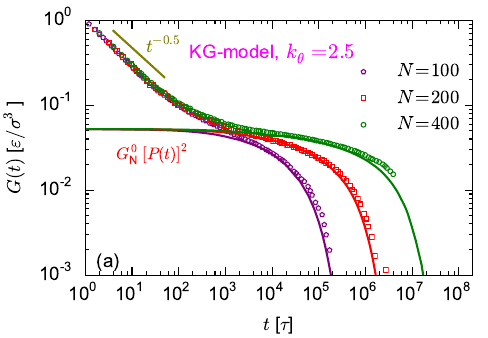}
    \end{subfigure}

    \begin{subfigure}{0.45\textwidth} 
        \includegraphics[width=1.0\textwidth]{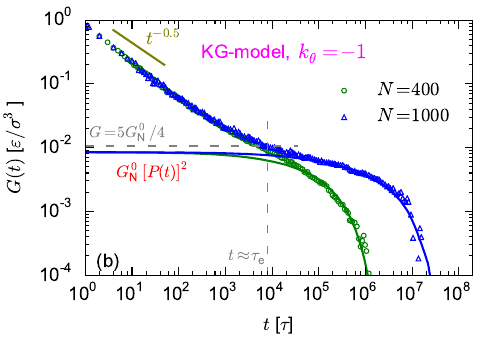}
    \end{subfigure}
  
    \begin{subfigure}{0.45\textwidth} 
        \includegraphics[width=1.0\textwidth]{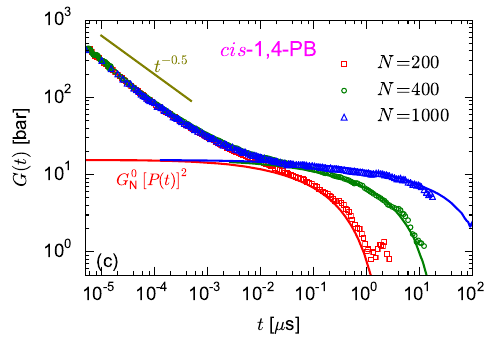}
    \end{subfigure}
\caption{Comparison between $G(t)$ and $G^0_\text{N}[P(t)]^2$ for three different polymer models.
Panel (a) shows the results for the bead–spring model with bending stiffness $k_\theta = 2.5$, where $G^0_\text{N} = 0.052\,\varepsilon/\sigma^3$, corresponding to $N_\text{e} = 13$.
Panel (b) presents the results for the bead–spring model with $k_\theta = -1$, for which $G^0_\text{N} = 0.0085\,\varepsilon/\sigma^3$, corresponding to $N_\text{e} = 80$.
In this panel, an estimate of the entanglement time is also shown, obtained from the condition $G(\tau_\text{e}) = \rho k_\text{B} T / N_\text{e} = 5 G^0_\text{N}/4$.
Panel (c) shows the relaxation functions for a chemistry-specific coarse-grained model of \emph{cis}-PB at $413$ K, where $G^0_\text{N} = 15.5\,\text{bar}$ corresponding to $N_\text{e} = 28$ monomers.}
    \label{Fig:double-2} 
\end{figure}

We continue by examining the validity of the relation $G(t) = G^0_\text{N} [P(t)]^2$ for the other systems studied (i.e., those aside from the Kremer–Grest model with $k_\theta = 1.5$). 
\cref{Fig:double-2}a shows the $G(t)$ curves for Kremer–Grest models with $k_\theta = 2.5$ and chain lengths $N = 100$, $200$, and $400$. In this panel, the $G(t)$ curves are compared with $G^0_\text{N}[P(t)]^2$ with $G^0_\text{N} = 0.052\, \varepsilon/\sigma^3$ corresponding to $N_\text{e} = 13$.
It can be observed that $G_\text{N}^0 [P(t)]^{2}$ describes the behavior of $G(t)$ well at times longer than the initial Rouse regime. 
For this stiffer model, the entanglement length obtained from the double reptation model, $N_\text{e} = 13$, is close to the value $N_{\text{e,ppa}} = 15$ calculated from primitive path analysis~\cite{everaers2020kremer}.

Panel (b) of \cref{Fig:double-2} shows the $G(t)$ data for the bead-spring model with $k_\theta = -1$ and chain lengths $N = 400$ and $1000$. For this model as well, the  $G(t) = G_\text{N}^0 [P(t)]^{2}$ with $G_\text{N}^0 = 0.0085\,\varepsilon/\sigma^{3}$ corresponding to $N_\text{e} = 80$, provides a good description of the $G(t)$ data.
This value of $N_\text{e}$ deviates from the estimate $N_{\text{e,ppa}} = 120$~\cite{everaers2020kremer}.

For the model with $k_\theta = -1$, similar to the case of $k_\theta = 1.5$, we also carry out the consistency check described above to further assess the entanglement lengths estimated from double-reptation relation and primitive-path analysis.
From the short-time Rouse regime of the measured $G(t)$ curves, we determine the monomeric time to be $\tau_0 = (1.5 \pm 0.2)\,\tau$.
With this value of $\tau_0$, we perform the self-consistency check described above.
Based on the $G(t)$ data for $N = 1000$, taking $N_\mathrm{e} = 80$ gives an entanglement time $\tau_\mathrm{e} \approx 8000\,\tau$ (see \cref{Fig:double-2}b). This value is relatively close to the Rouse time of a subchain of length $N_\mathrm{e} = 80$, namely $\tau_0 N_\mathrm{e}^2 = 9600\,\tau$.
In contrast, using $N_{\mathrm{e,ppa}} = 120$ yields $\tau_\mathrm{e} \approx 45000\,\tau$, which is significantly larger than the corresponding Rouse time, $\tau_0 N_{\mathrm{e,ppa}}^2 = 21600\,\tau$.

\begin{table}[h]
\centering
\begin{tabular}{|c|c|c|}
\hline
$k_{\theta}$ & $N_\text{e}$ & $N_\text{e,ppa}$  \\ \hline
$-1$ & $80$   & $120$~\cite{everaers2020kremer}    \\
$0$ &  $52$  &  $87$~\cite{hoy2009topological,moreira2015direct}, $78$~\cite{everaers2020kremer}   \\
$1.5$ & $21$  &  $28$~\cite{hsu2016static}, $27$~\cite{everaers2020kremer}  \\
$2.5$ & $13$  &  $15$~\cite{everaers2020kremer}    \\ \hline
\end{tabular}
\caption{Entanglement lengths $N_\text{e}$ corresponding to the plateau modulus, as obtained from the relation $G(t) = G^0_\text{N} [P(t)]^2$, for bead–spring chains of varying stiffness specified by $k_\theta$. Entanglement lengths determined from primitive-path analysis, $N_\text{e,ppa}$, are also included. The error bars on the $N_\text{e}$ values were estimated to be around $10\%$  by (1) dividing the simulation trajectory into three blocks and computing $G(t)$ for each block, thus assessing the variation of $G(t)$ between blocks, and (2) varying the plateau modulus (which directly corresponds to $N_\text{e}$) in the relation $G(t) = G^0_{\text{N}}[P(t)]^2$ and evaluating the quality of the resulting fits.}
\label{Tab:Ne}
\end{table}

The $N_\text{e}$ values extracted from the relation $G(t) = G^0_\text{N}[P(t)]^2$ for Kremer–Grest chains of varying stiffness are summarized in \cref{Tab:Ne}. The result for the standard Kremer–Grest model with $k_{\theta} = 0$ from our previous work~\cite{behbahani2024relaxation} is also included. For comparison, the entanglement lengths calculated via primitive path analysis~\cite{hoy2009topological,moreira2015direct,hsu2016static,svaneborg2020characteristic,everaers2020kremer} are reported in the table as well.
The results from double reptation and primitive path analysis behave qualitatively similarly, and both decrease with increasing $k_{\theta}$ and thus increasing chain stiffness.
However, it seems that $N_{\text{e,ppa}}$ is systematically larger than the value from double reptation, meaning that primitive path analysis slightly underestimates the plateau modulus, as inferred from the double reptation relation.
For stiff chains, the results of the primitive path analysis are close to those obtained from the double reptation relation; however, for flexible chains, the results have a relatively larger difference.
As noted above, based on slip-spring simulations, Likhtman et al.~\cite{likhtman2007linear} also reported that primitive path analysis underestimates the plateau modulus.

Finally, panel (c) of \cref{Fig:double-2} compares the $G(t)$ data for \emph{cis}-PB with the prediction $G^0_\text{N}[P(t)]^2$, using $G^0_\text{N} = 15.5\,\text{bar}$, which corresponds to an entanglement length of $N_\text{e} = 28$ monomers ($T = 413\,\text{K}$).
Results are shown for chain lengths $N = 200$, $400$, and $1000$ monomers, and good agreement between the measured $G(t)$ curves and $G^0_\text{N}[P(t)]^2$ is observed.
The entanglement length $N_\text{e} = 28$ corresponds to an entanglement molecular weight of $M_\text{e} = 1512\, \text{g/mol}$, which is comparable to an available experimental value of $M_\text{e} = 1800\,\text{g/mol}$ for a 1,4-rich PB sample~\cite{fetters1994connection}, indicating a reasonable fidelity of the employed coarse-grained model~\cite{behbahani2021dynamics} in this regard.

Overall, in this section, 
we observed very good agreement between $G(t)$ and $G^0_\text{N}[P(t)]^2$ for all monodisperse polymer melts studied.

\subsection{Comparison of simulation results with theoretical expressions}
\label{sec:theo}

The relation $G(t) = G^0_\text{N}[P(t)]^2$ simplifies the modeling of $G(t)$ for monodisperse samples. Based on this relation, an accurate analytical expression for $P(t)$ is sufficient to predict the long-time behavior of $G(t)$ in monodisperse melts. In this section, we therefore compare the simulation data for $P(t)$ with analytical expressions. We also compare the measured $G(t)$ data with theoretical predictions based on double reptation. 

As noted in \cref{sec:intro}, within the tube model, $P(t)$ is equivalent to the surviving tube fraction, $\mu(t)$. This equivalence, however, neglects the short-time decay of $P(t)$ due to motion within the tube. In what follows, we compare $P(t)$ with an analytical expression for $\mu(t)$ that accounts for contour-length fluctuations (CLF) and reptation, calculated within the fixed-tube model.
For short chains, particularly those in the crossover regime between Rouse and entangled dynamics, $\mu(t)$ calculated within a fixed-tube framework is expected to show limitations when used to model $P(t)$. In such systems, the Rouse regime, which is absent from $\mu(t)$, contributes significantly to the decay of $P(t)$. Furthermore,
the characteristic behavior of $P(t)$ may differ from the fixed-tube prediction~\cite{behbahani2024relaxation}.
With increasing chain length in a monodisperse melt, the contribution of the Rouse regime decreases, and the characteristic behavior of $P(t)$ approaches that predicted by the fixed-tube model~\cite{mcleish2002tube}.

For describing $P(t)$, we use the following analytical expression proposed by Likhtman and McLeish~\cite{likhtman2002quantitative}, for the evolution of $\mu(t)$ due to contour-length fluctuations at short times and reptation at longer times:
\begin{equation}
\begin{aligned}
        \mu(t) &= \int_{\varepsilon^*(Z,\tau_\text{e})}^{\infty} \frac{0.306}{Z\tau_\text{e}^{1/4}\varepsilon^{5/4}}\exp(\varepsilon t)\mathrm{d}\varepsilon \\ 
        &+ \frac{8\Tilde{G}_\text{f}(Z)}{\pi^2} \sum_{p, \text{odd}}^{p^*(Z)} \frac{1}{p^2}\exp(-\frac{p^2 t}{\tau_\text{df}(Z,\tau_\text{e})}). 
        \label{Eq:LM}
\end{aligned}
\end{equation}
The first term of \cref{Eq:LM} represents the contribution from CLF and dominates up to times on the order of the Rouse time $\tau_\text{R}$. This term is responsible for the scaling behavior $(1 - \mu(t)) \sim t^{1/4}$. The second term corresponds to the contribution from reptation, which becomes dominant at times longer than $\tau_\text{R}$.
The time $\tau_\text{df}(Z)$ denotes the disentanglement time in the presence of CLF, and $\Tilde{G}_\text{f}(Z)$ is a dimensionless plateau modulus. Explicit expressions for these quantities, $\varepsilon^*$, and $p^*$ can be found in ~\textcite{likhtman2002quantitative}.

\begin{figure}[htb!]
    \centering

\begin{subfigure}{0.45\textwidth} 
        \includegraphics[width=1.0\textwidth]{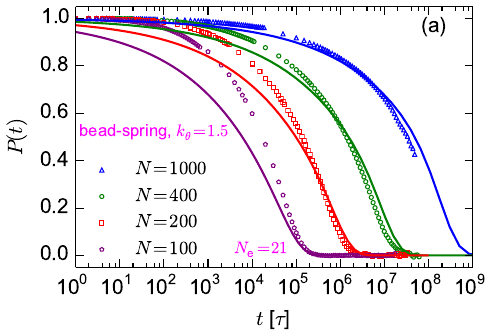}
    \end{subfigure}

    \begin{subfigure}{0.45\textwidth} 
        \includegraphics[width=1.0\textwidth]{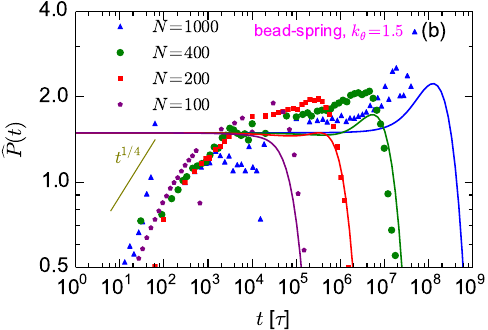}
    \end{subfigure}
  
\caption{(a) The $P(t)$ data for various chain lengths of the semi-flexible bead–spring model with $k_\theta = 1.5$ (symbols), shown together with the $\mu(t)$ in the Likhtman–McLeish model, \cref{Eq:LM} (solid lines), using $N_\text{e} = 21$ and $\tau_\text{e} = \tau_0 N_\text{e}^2 = 2.1 N_\text{e}^2$. The model inputs were determined independently and were not treated as free fitting parameters.
(b) The same data presented in the format adopted by Likhtman and McLeish~\cite{likhtman2002quantitative}, where $\widehat{P}(t) = -4Z\tau_\text{e}^{1/4} t^{3/4} \partial P/\partial t$.}
    \label{Fig:pt} 
\end{figure}

The comparison of simulation data for $P(t)$ with \cref{Eq:LM} is performed for the semiflexible Kremer–Grest model with $k_\theta = 1.5$, as a wider range of chain lengths has been simulated for this system.
The quantities $Z = N/N_\text{e}$ and $\tau_\text{e}$ are the only input parameters in \cref{Eq:LM}.
For the bead-spring model with $k_\theta = 1.5$, we set $N_\text{e} = 21$, the value obtained from double reptation model (see \cref{Fig:Gt-Pt-K1.5} and \cref{Tab:Ne}). We then calculate $\tau_\text{e}$ using $\tau_\text{e} = \tau_0 N_\text{e}^2$, with $\tau_0 = 2.1\,\tau$ independently determined from the Rouse regime of the measured $G(t)$ functions, as discussed above in the context of \cref{Fig:Gt-Pt-K1.5} (see also \cref{Tab:tau_0}).
\cref{Fig:pt} shows the measured $P(t)$ curves (symbols) for various chain lengths of the bead–spring model with $k_\theta = 1.5$ together with the predictions of \cref{Eq:LM} (lines).
The data are presented in two formats. Panel (a) of \cref{Fig:pt} shows $P(t)$ vs.\ $t$, while panel (b) presents $\widehat{P}(t) = -4Z\tau_\text{e}^{1/4} t^{3/4} \partial P/\partial t$ vs. $t$, following the representation adopted by Likhtman and McLeish~\cite{likhtman2002quantitative}.
In this latter representation, the CLF regime, where $(1 - P(t)) \sim t^{0.25}$, appears as a plateau, while the long-time reptation regime appears as a peak.

\cref{Eq:LM} poorly describes the $P(t)$ data for the shortest chain studied, but the agreement with the $P(t)$ data improves as the chain length increases; nevertheless, even for long chains, differences between the prediction and the simulation data persist at long times.
A major reason for the inadequacy of \cref{Eq:LM} in describing the $P(t)$ of short chains is the neglect of the initial short-time Rouse regime.
At short times, before the chains feel entanglements, $P(t)$  exhibit a Rouse regime in which $(1 - P(t)) \sim t^{0.5}$ or $\widehat{P}(t) \sim t^{0.25}$ (see panel (b) of \cref{Fig:pt})~\cite{behbahani2024relaxation}. For short, mildly entangled chains, the contribution of this Rouse regime to the decay of $P(t)$ is significant; however, for long chains, the contribution of this regime becomes negligible.
Previous studies~\cite{hou2010stress,ghanta2025modeling} have proposed modifications to \cref{Eq:LM}, particularly to improve its short-time behavior. These modified equations introduce one or more additional fitting parameters that can be adjusted to improve the quality of the fit, which, as mentioned, is most important for shorter chains.
In the current work, we stay with the basic model and, as noted above, determine its input parameters independently.
Indeed, the quality of the fit can be slightly improved at long times if the inputs to \cref{Eq:LM} are treated as free fitting parameters, for example, by reducing $\tau_0$ relative to the value determined from the short-time regime of $G(t)$. With the current input parameters, the theoretical prediction is slightly higher than the measured curves at long times. It is worth noting that this long-time effect is slightly more pronounced for the stiffest chains with $k_\theta = 2.5$ and weaker for the system with $k_\theta = -1$ (data not shown).

\begin{table}[h]
\centering
\begin{tabular}{|c|c|c|}
\hline
model & $\tau_0$  & $b$   \\ \hline
KG, $k_{\theta} = -1$ &  $1.5\,\tau$ & $1.25\,\sigma$    \\
KG, $k_{\theta} = 1.5$ & $2.1\,\tau$   & $1.65\,\sigma$ \\
KG, $k_{\theta} = 2.5 $ & $3.0\,\tau$   &$1.99\,\sigma$    \\ 
\emph{cis}-PB ($T = 413$ K) & $1.3$ ps   & $0.69$ nm       \\
\hline
\end{tabular}
\caption{The values of the monomeric time  $\tau_0[ = \zeta b^2/(3\pi^2)]$  and statistical segment length for the polymer melts studied.
$\tau_0$ is determined from the short-time Rouse regime of the $G(t)$ curves; see \cref{Fig:Gt-Pt-K1.5}a and the corresponding discussion. The statistical segment length $b$ is determined from the large-scale values of internal chain distances.}
\label{Tab:tau_0}
\end{table}

Having compared the $P(t)$ data with the analytical expression, we now compare the $G(t)$ data to the analytical prediction given by the double reptation (and dynamic tube dilation) model $G(t)=G^0_\text{N}[P(t)]^2$.
Based on the results of \cref{sec:double-rep}, the long-time behavior of $G(t)$ is described by $G^0_\text{N}[P(t)]^2$. Using the analytical relation for $\mu(t)$ (\cref{Eq:LM}) to describe $P(t)$, we then have $G(t) = G^0_\text{N}[\mu(t)]^2$, which is the original relation proposed by Marrucci~\cite{marrucci1985relaxation} and des Cloizeaux~\cite{des1988double,des1990relaxation,des1993polymer}.
The contribution of short-time processes should also be included to fully describe $G(t)$ from the Rouse regime up to the terminal time. Here, these contributions are taken from the Likhtman–McLeish model~\cite{likhtman2002quantitative}.
The resulting equation is:
\begin{equation}
 \begin{aligned}
         G(t) &=  G^0_\text{N} [\mu(t)]^2 + \frac{\rho k_\text{B}T}{N}  \sum_{p=Z}^{N} 
           \exp(-\frac{2p^2 t}{\tau_\text{R}})\\
         &+ \frac{\rho k_\text{B}T}{5N} \sum_{p=1}^{Z-1} 
           \exp(-\frac{p^2 t}{\tau_\text{R}})  
    \label{Eq:Gt-LM}  
 \end{aligned}
\end{equation}
where $ G^0_\text{N} = (4\rho k_\text{B}T)/(5N_\text{e})$ and $\tau_\text{R} = \tau_0 N^2$ is the Rouse time of the chain. 
The first term is, as discussed above, the contribution from double reptation; the second term represents the contribution from the initial Rouse regime; and the third term accounts for the relaxation of longitudinal modes, which leads to a redistribution of segments along the chain.
In the original Likhtman–McLeish model~\cite{likhtman2002quantitative}, the first term takes the form $G^0_\text{N}\mu(t) R(t,c_\nu)$, where, as noted in \cref{sec:intro}, $R(t,c_\nu)$ is calculated based on the Rouse–tube model of CR, with $c_\nu$ being a fitting parameter which significantly controls the decay rate of $R(t,c_\nu)$. Replacing $G^0_\text{N}\mu(t) R(t,c_\nu)$ with $G^0_\text{N}[\mu(t)]^2$ in \cref{Eq:Gt-LM} is a natural consequence of the relation $G(t) = G^0_\text{N}[P(t)]^2$, which was examined in \cref{sec:double-rep}.
As noted in \cref{sec:intro}, $G^0_\text{N}\mu(t) R(t,c_\nu)$ does not yield $G^0_\text{N}[P(t)]^2$ because $R(t,c_\nu)$ can be significancy different from $\mu(t)$.
For $c_\nu \approx 1$,
the function $R(t,c_\nu)$ 
is close to $\mu(t)$ for $t<\tau_\text{R}$, but it has long-time tail beyond $\tau_\text{R}$ and deviates from $\mu(t)$~\cite{likhtman2002quantitative}.
In some previous studies $R(t,c_\nu)$ has been approximated by $\mu(t)$, and \cref{Eq:Gt-LM} has been used to describe $G(t)$~\cite{hou2010stress,li2021dynamics,ghanta2025modeling}.
However, the results of \cref{sec:double-rep} regarding the examination of the double-reptation relation provide a basis for the form $G^0_\text{N}[\mu(t)]^2$, rather than viewing it as an approximation to the result obtained from the Rouse–tube model of CR.

\begin{figure*}[htb!]
    \centering
\begin{subfigure}{0.45\textwidth} 
        \includegraphics[width=1.0\textwidth]{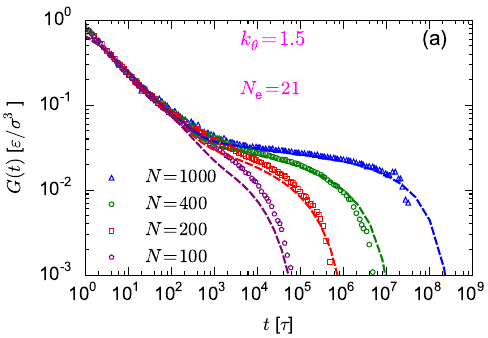}
\end{subfigure}
\begin{subfigure}{0.45\textwidth} 
        \includegraphics[width=1.0\textwidth]{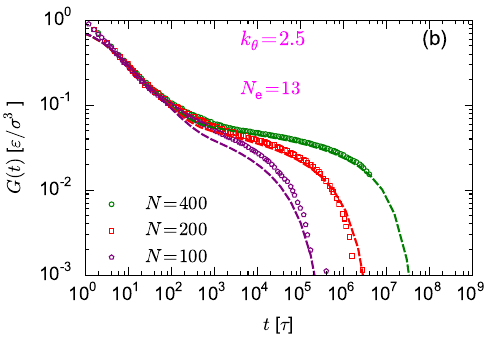}
\end{subfigure}

\begin{subfigure}{0.45\textwidth} 
        \includegraphics[width=1.0\textwidth]{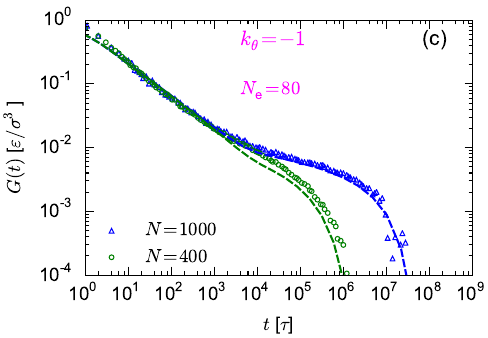}
\end{subfigure}
\begin{subfigure}{0.45\textwidth} 
        \includegraphics[width=1.0\textwidth]{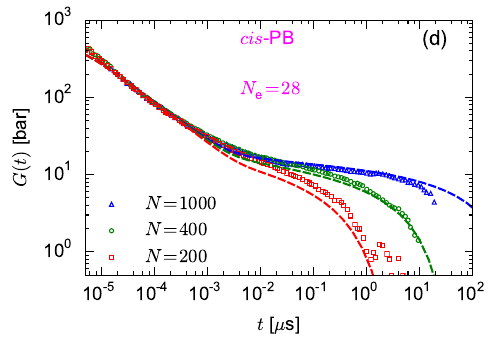}
\end{subfigure}

\caption{The $G(t)$ curves for the polymer melts studied (symbols), shown together with the predictions of \cref{Eq:Gt-LM} (dashed lines). Panels (a), (b), and (c) show the results for the bead-spring models with $k_{\theta} = 1.5$, $k_{\theta} = 2.5$, and $k_{\theta} = -1$, respectively. Panel (d) shows the results for \emph{cis}-PB at 413 K. \cref{Eq:Gt-LM} uses $N_\text{e}$ and $\tau_\text{0}$ as explicit inputs. The values of $N_\text{e}$ are taken from the double reptation analysis reported in \cref{Tab:Ne}, and the monomeric times $\tau_0$ are obtained from the Rouse regimes of the $G(t)$ curves, as reported in \cref{Tab:tau_0}.}
    \label{Fig:gt-lm} 
\end{figure*}

Similar to \cref{Eq:LM} for $\mu(t)$, the input parameters of \cref{Eq:Gt-LM} are $\tau_0$ and $N_\text{e}$, which we determine independently from the double reptation analysis and the initial Rouse regime of $G(t)$, respectively. The values of $N_\text{e}$ and $\tau_0$ for the different models are reported in \cref{Tab:Ne} and \cref{Tab:tau_0}, respectively.
\cref{Fig:gt-lm} shows the measured $G(t)$ curves (symbols) for the polymer melts studied, together with the predictions of \cref{Eq:Gt-LM} (dashed lines).
The quality of \cref{Eq:Gt-LM} in describing the measured $G(t)$ data follows the same trend observed in \cref{Fig:pt} for the performance of \cref{Eq:LM} in describing $P(t)$, as expected from the relation $G(t) = G^0_\text{N}[P(t)]^2$.
The function $P(t)$ for short chains with $Z \approx 5$, is not well described by \cref{Eq:LM}, leading to large deviations of \cref{Eq:Gt-LM} from the measured $G(t)$. However, with increasing chain length, the agreement between the theoretical predictions and the simulation results improves, and overall the theoretical predictions show fair agreement with the simulation data.

\section{Summary}
\label{sec:summary}

We investigated stress relaxation in entangled polymer melts by comparing the shear stress–relaxation modulus, $G(t)$, with the end-to-end vector autocorrelation function, $P(t)$, in several types of strictly monodisperse, entangled linear polymer melts using molecular dynamics simulations.
The investigation includes three bead–spring models with varying chain stiffness, as well as a chemistry-specific coarse-grained model of \emph{cis}-1,4-polybutadiene.
For each model, several chain lengths were simulated, spanning a range of approximately $N/N_\text{e} = 5$–$50$ entanglements per chain.

For all of the studied samples, it was observed that 
$G^0_\text{N}[P(t)]^2$ very well describes $G(t)$ beyond the short-time Rouse regime.
Here, the chain-length-independent prefactor $G^0_\text{N}$ is the plateau modulus, which is related to the entanglement length by $G^0_\text{N} = 4\rho k_\text{B}T/(5N_\text{e})$.
This observation is consistent with the prediction of the double reptation model, and also the dynamic tube dilation model for the relaxation of polymer melts, although the two models are based on different physical pictures.
The dynamic tube dilation model is based on the assumption that the portion of a chain that escapes the tube behaves as a solvent, leading to an increase in the tube diameter.   
The double reptation model is based on a particularly simple network-based picture of stress relaxation in polymer melts, in which an entangled polymer melt is viewed as a temporary network where each node is formed by an entanglement between two polymer chains, and the stress relaxation modulus is proportional to the number of original nodes that remain at time $t$.

The relation $G(t) = G^0_{\text{N}} [P(t)]^2$, can serve as a basis for determining the plateau modulus and the corresponding entanglement length. For this purpose, high-quality $G(t)$ and $P(t)$ data are required for a few chain lengths. The plateau modulus $G^0_{\mathrm{N}}$ is then determined as a chain-length-independent prefactor that satisfies this relation.
For the Kremer-Grest models of different stiffnesses, we compared the so-obtained values of the entanglement lengths with those obtained from primitive path analysis. The results from both methods show similar trends, with both decreasing as chain stiffness increases. 
However, it seems that $N_{\text{e,ppa}}$, calculated from primitive path analysis, is systematically larger than the value from double reptation, meaning that primitive path analysis slightly underestimates the plateau modulus.
For stiff chains, the results of the primitive path analysis are close to those obtained from the double reptation relation; however, for flexible chains, the differences are relatively larger.

Based on the relation $G(t) = G^0_{\text{N}} [P(t)]^2$, an accurate analytical expression for $P(t)$ is sufficient to model the long-time behavior of $G(t)$. To describe $P(t)$, we used the analytical expression proposed by Likhtman and McLeish that incorporates contour-length fluctuations and reptation. The resulting prediction for $G(t)$ is then compared with the simulation data. Fair agreement between the simulation results and theoretical predictions is observed.

In the present work, we investigated the viscoelastic response of monodisperse linear polymer melts. For these systems, our simulation results are in good agreement with the predictions of the double reptation and dynamic tube dilation models.
A further examination of the prediction of these models can be performed by analyzing stress relaxation in samples with chain-length dispersity.
This is particularly important because previous experimental studies~\cite{watanabe2004viscoelastic,watanabe2004test,van2012effective} have reported shortcomings of these models, in their basic forms, in describing the viscoelastic properties of mixtures of long and short polymer chains.
The investigation of such systems will be the subject of a future study.

\subsection*{Acknowledgements}
We thank Friederike Schmid for careful reading of the manuscript and for useful comments.
This research was supported by the German Science Foundation
(DFG) via SFB TRR 146 (Grant number 233630050, project C1). 
The simulations were carried out on the high
performance computing center MOGON at JGU Mainz.

\end{document}